\documentclass[acus]{JAC2000}

\usepackage{graphicx}
\newcommand{\im}{\mbox{Im}}
\newcommand{\re}{\mbox{Re}}

\begin{document}
\title{\flushright{T19}\\[15pt] \centering 
DISPERSIVE TECHNIQUES FOR $\alpha_s$, $R_{had}$
AND INSTABILITY OF THE PERTURBATIVE VACUUM}

\author{Y. Srivastava, Dipartimento di Fisica dell'Universit\`a and INFN, Perugia, 
Italy and\\
Physics Departement, Northeastern University, Boston, MASS, USA\\
S. Pacetti, Dipartimento di Fisica dell'Universit\`a and INFN, Perugia, Italy\\
G. Pancheri, INFN Laboratori Nazionali di Frascati, Italy\\
A. Widom, Northeastern University, Boston, MASS, USA}

\maketitle

\begin{abstract}

 Recent dispersive techniques developed by us are applied to
discuss three problems: 1. A long standing discrepan-cy between the
measurements of $R(s)$ for $\sqrt{s}\ = (5\div 7.5)GeV$
by Crystal Ball and MARK I has been analyzed and its consequences
analyzed for the number of contributing quarks. 2. Noting that the 
perturbative $\alpha_s$ has the wrong analyticity, analytic models 
consistent with asymptotic freedom (AF) and confinement have been 
constructed and applied to discuss $\tau$ decay. 3. It is shown that 
AF leads to a wrong sign for $\im\big(\alpha(s)\big)$ which 
signals an instability of the perturbative QCD vacuum. 

\end{abstract}

\section{ The dispersive method}

Recently, we have developed accurate numerical schemes to handle 
the ``inverse'' problem, that is how to retrieve the imaginary part
of a physical quantity (e.g., a cross section, or a form factor
in the time-like region) using data for its real or the reactive part 
(usually in the space like region)\cite{R1,R2}. 
\begin{figure*}[t]\vspace{-.0cm}
\includegraphics*[width=82.5mm,height=82.5mm]{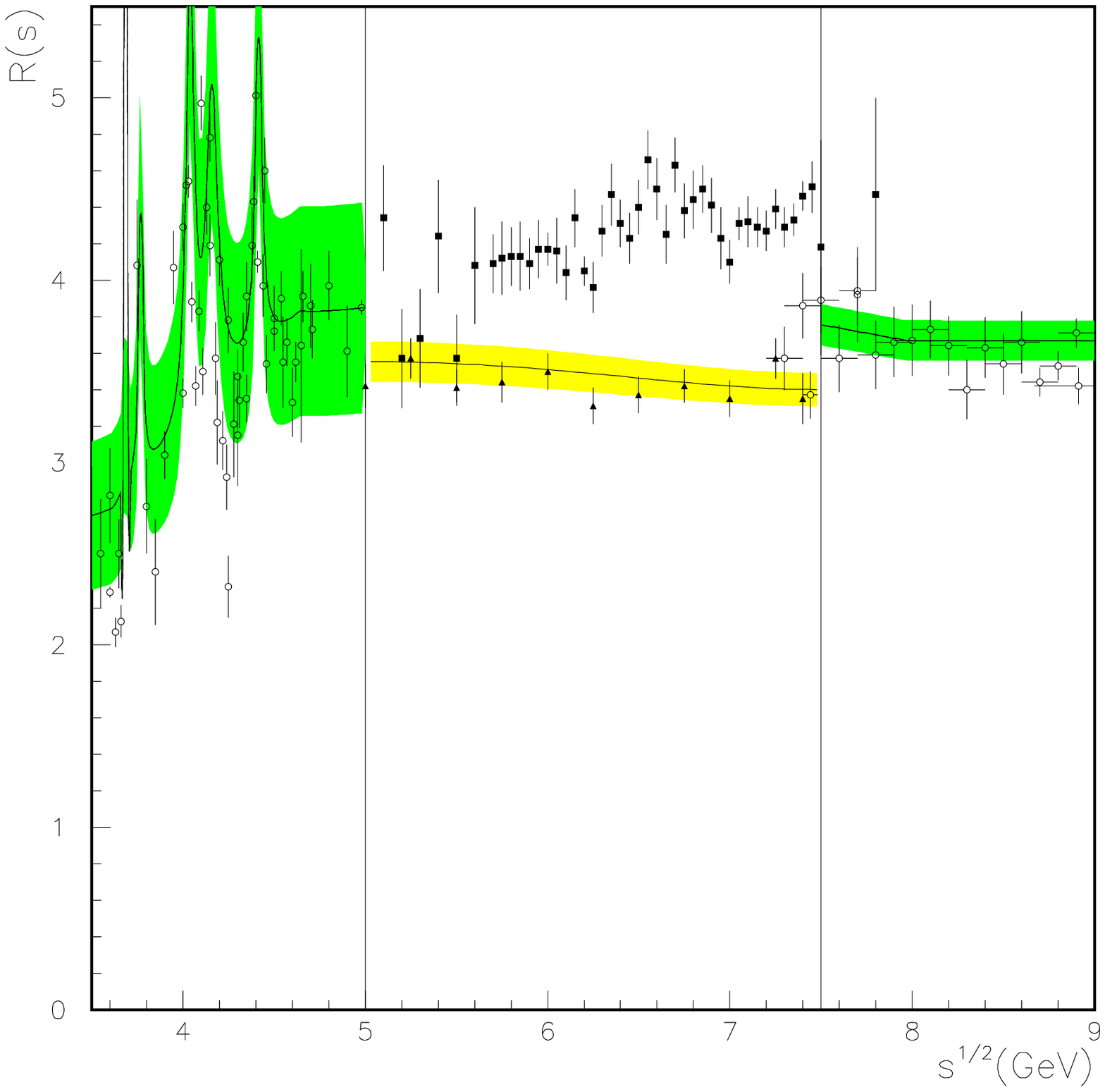}
\includegraphics*[width=82.5mm,height=82.5mm]{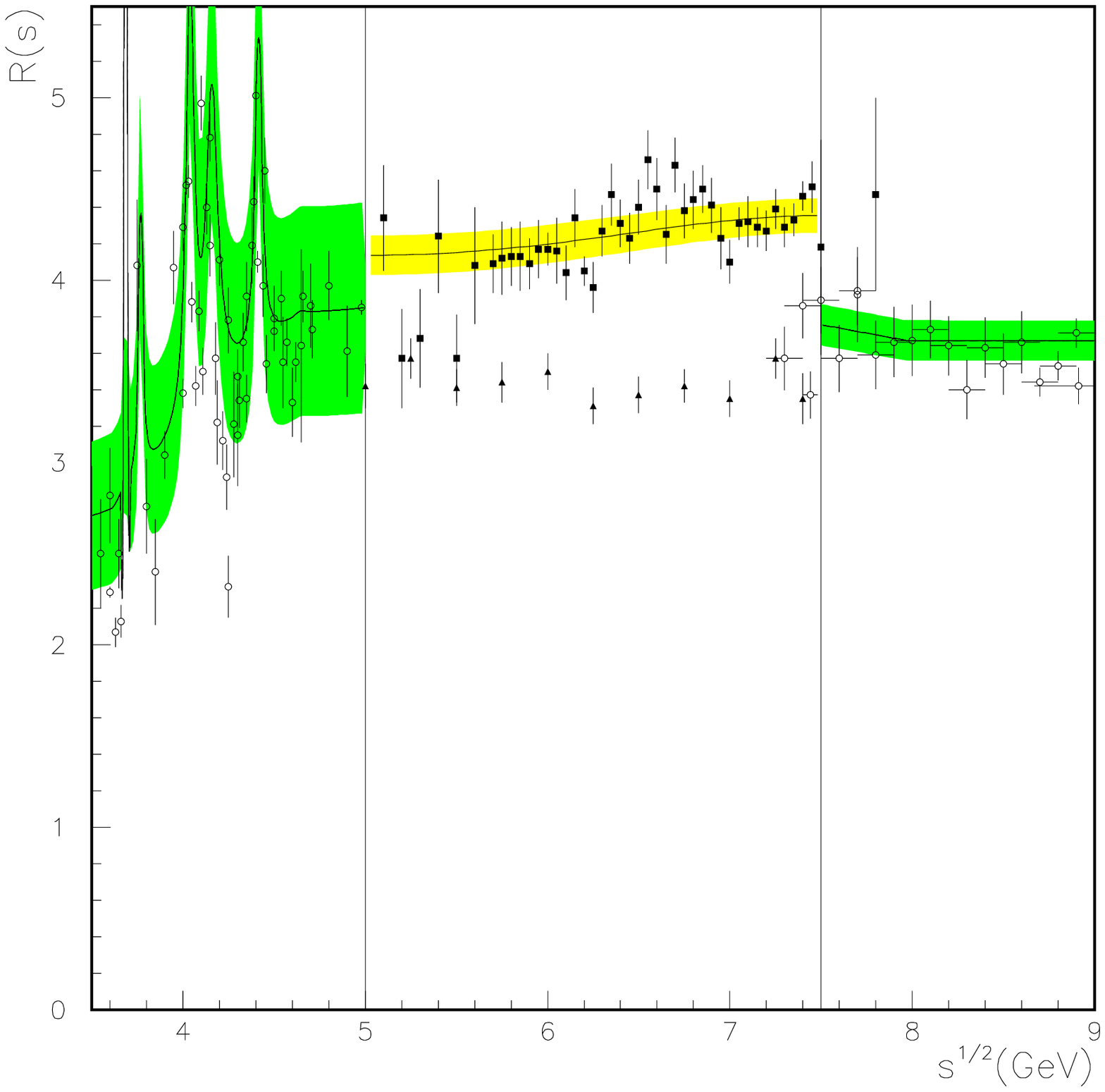}
\begin{minipage}[t]{7.5cm}\vspace{-.5cm}
Figure 1: Values of $R(s)$ (light gray band) obtained from the integral 
equation with five spin 1/2 quarks in the interval $[5\;GeV,7.5\;GeV]$ of
$s$.
\end{minipage}
\ \hspace{1cm} \
\begin{minipage}[t]{7.5cm}\vspace{-.5cm}
Figure 2: Values of $R(s)$ (light gray band) obtained from the integral 
equation with five spin 1/2 quarks plus one scalar quark.
\end{minipage}
\end{figure*}
It has been reviewed at this conference by S. Pacetti (see his 
contribution T20), and its applications to the nucleon form factor  
have been discussed. Here these techniques have been applied to three
following problems.

\section{ Determination of $R(s)$ for $\sqrt{s}\ = 
(5\div 7.5)GeV$}

The purpose of this analysis is to once again focus on a long standing
discrepancy between data from Crystal Ball \cite{C1} and MARK I \cite{C2}
in the energy region $\sqrt{s}\ = (5\div 7.5)GeV$. To arrive at
a model independent answer to this problem we have applied the 
following procedure (for details we refer the reader to \cite{Q1}). 
We write a dispersion relation for the  derivative of the 
polarization tensor $\Pi(t)$ whose imaginary part is proportional to 
$R(s)$. As input, we use all available data for $R(s)$ from outside 
the disputed region and for its  asymptotic part contributions from
the five light quark flavors. We then solve the integral equation for
$R(s)$ in the disputed region. As shown in Fig.(1), very good agreement
is found with the CB data. This implies that some additional contribution 
to the asymptotic behavior is mandatory to reproduce the MARK I data. 
As shown in Fig.(2), we find that a low mass, spin zero quark of 
charge $(-1/3)$ is quite adequate for this purpose. 

It is worthwhile to point out that it is not inconceivable for the CB data
to be consistent with the MARK I data given their different selection
criteria for what constituted $R(s)$. If one inquires into the selection 
criteria for $R(s)$ in the CB data, one is struck by the remarkable fact 
that in it only two jet ($q\bar{q}$) events were included. More precisely, 
all events with more than 20\% imbalance in energy between forward-backward, 
left-right and up-down hemispheres were discarded. Were there 
any decays from beyond that of the simple ($q\bar{q}$) type, they were 
not counted. MARK I had imposed no such restriction.

Given the importance consequences for the standard model that it entails, 
we stress upon the necessity of an independent measurement of $R$ in 
this region. One possibility is through the radiative technique at 
$B\bar{B}$ machines\cite{Q1}.     

\section{ Analytic models for $\alpha_s$ with application to
$\tau$ decay   }

Several authors have noted that the perturbative $\alpha_s(s)$ has 
the wrong analyticity\cite {RU1,RU2}. E.g., the 1-loop AF
formula
$$
\alpha_{1-loop}(s) = \left({{1}\over{b}}\right)
{{1}\over{\ln(-s/\Lambda^2)}},
\eqno(1)
$$ 
has a pole at space-like value $s=-\Lambda^2$. Higher loops suffer
from the same disease. Of course, analyticity derived from unitarity
forbids any singularity for space like $s$ ($<0$).

Several cures for the above have been suggested. For example, in 
\cite{RU1} the imaginary part for $\alpha_s$ computed from AF
$$
\im\big(\alpha_1(s)\big) = \left({{1}\over{b}}\right){{\pi}\over
{\ln(s/\Lambda^2)^2 + \pi^2}} \vartheta(s),\eqno(2)
$$
is used in an unsubtracted dispersion relation (which converges
thanks to the behavior $1/(\ln(s/\Lambda^2))^2$ in the asymptotic
region) to compute the real part. Thus, one has
$$
\alpha_1(s) = {{1}\over{b}}\left[ {{1}\over{\ln(-s/\Lambda^2)}}
+ {{\Lambda^2}\over{\Lambda^2 + s}}\right].\eqno(3)
$$
The second term cancels the unwanted pole. This procedure has been
generalized upto 3-loops. An added curiosity: $\alpha_I(0)\ =\ (1/b)$,
is finite and universal to all loops.\par
So why not stop here ? The lacuna is that $\alpha_I(s)$ is too tame;
has not enough ``oomph'' to produce all what QCD is advertised to
possess. That is, to obtain, confinement of quarks and glue, infinite 
number of Regge trajectories, etc. etc., one would then have to 
add - in an ad hoc fashion - a confining potential to produce a 
reasonable hadronic spectrum. E.g., on the lattice, where the Wilson 
area law is imposed automatically and thereby confinement, so there 
$\alpha(0) \rightarrow$ constant or even zero. Parenthetically, the 
embarrassment on the lattice is that the same prescription leads to 
a linear potential $V(r)\rightarrow r$ rather than $(1/r)$ for QED.\par
A different analytic model is due to Nesterenko\cite{RU2}, where the
AF pole is eliminated multiplicatively:
$$
\alpha_{II}(s) = {{1}\over{b}}\left[{{(1 + {{\Lambda^2}\over{s}})}\over
{\ln(-s/\Lambda^2)}}\right].\eqno(4)
$$
Here $\alpha_{II}(s)$ increases as $s$ goes to zero.\par
The problem with this model is that it is ``too singular'', so
that integrals (such as those needed for soft gluon summations
\cite{Pancheri1})
$$
\alpha_{av} = {{1}\over{s}}\int_0^s ds' \alpha(s'), \eqno(5)
$$    
are not convergent.\par
Taking our cue from the above, we have developed the following
general strategy to develop a whole class of models for an
analytic $\alpha$ - with 1 and 2 as special cases.\par
A simpler function to disperse is ($1/\alpha(s)$), which provides
much physical insight. Let us recall that the vacuum 
Coulomb potential is modified in the following way for charged 
particles in a medium
$$
{{\alpha_o}\over{r}}\ \ \ \rightarrow\ \ {{\alpha_o}\over{\epsilon r}},
\eqno(6)
$$ 
where $\epsilon$ is the dielectric constant of the material. Hence,
we shall consider
$$
\epsilon(s) = {{1}\over{\alpha(s)}}.\eqno(7)
$$
E.g., in the AF case, its imaginary part is much simpler, a constant
$$
\im\big(\epsilon(s)\big) = - \pi b \vartheta(s). \eqno(8)
$$ 
We write a subtracted dispersion relation for $\epsilon$
$$
\epsilon(s) = \epsilon(-\Lambda^2) + {{(s + \Lambda^2)}\over{\pi}}
\!\!\!\int_0^\infty\!\!\!\!\!\!\! {{ds'\ \im \big(\epsilon(s')\big)}\over
{(s' -s-i\delta)(s' + \Lambda^2)}}. \eqno(9)
$$
In Eq.(9), confinement is easily imposed. Here, confinement means 
$$
\epsilon(s=0)=0. \eqno(10)
$$
Using Eq.(8) as the asymptotic limit from AF for $\im\big(\epsilon(s)\big)$,
we have a general class of models consistent with AF and confinement
provided by [Fig.(3)]
$$
\im\big(\epsilon(s; p)\big) = - \pi b {{1}\over{1 + (\Lambda^2/s)^p}};
\hspace{.5cm} ( 0 < p \leq 1). \eqno(11) 
$$
The model in \cite{RU2} is obtained for $p=1$.\\
We have considered two applications: 1. $\tau$ decay, Fig.(4); 2. Hadronic 
transverse momentum distributions in W and Z decays. For 
$R_\tau^{hadronic}$, AF analysis requires a large $\Lambda 
\approx 850\;MeV$. For values of $p\approx (0.5\div 0.8)$, on the
other hand, we obtain a much more reasonable value for $\Lambda 
\approx 300\; MeV$ to get agreement with the data.\par
Regarding the  transverse momentum distribution broadening due to 
soft gluon summations, we reproduce previous phenomenological results
\cite{Pancheri2} for values of $p$ in the same range. 
Both these applications lead us to conclude that the
dispersive method is indeed capable of joining AF with confinement
in a suitable way.
\begin{figure*}[t]
\centering
\includegraphics*[width=110mm]{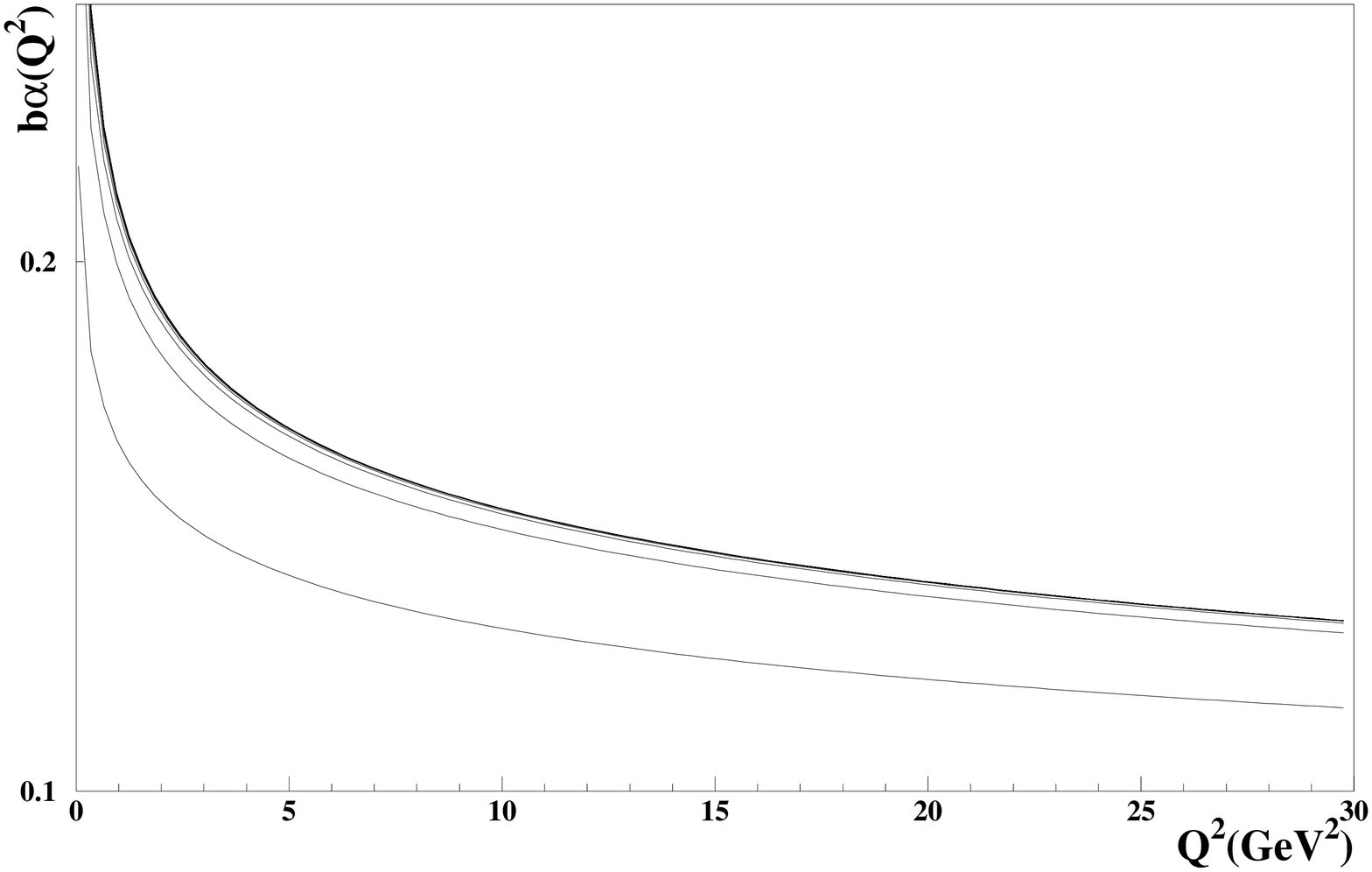}
\begin{minipage}[t]{10cm}\vspace{-.5cm}
Figure 3: Values of $\alpha(Q^2)$ space-like obtained by integrating the 
dispersion relation (9) with the imaginary part (11) for $p=1/5$ (lower curve),
$2/5,...,4$ (higher curve) ($\Lambda=100\;MeV $).
\end{minipage}
\includegraphics*[width=110mm]{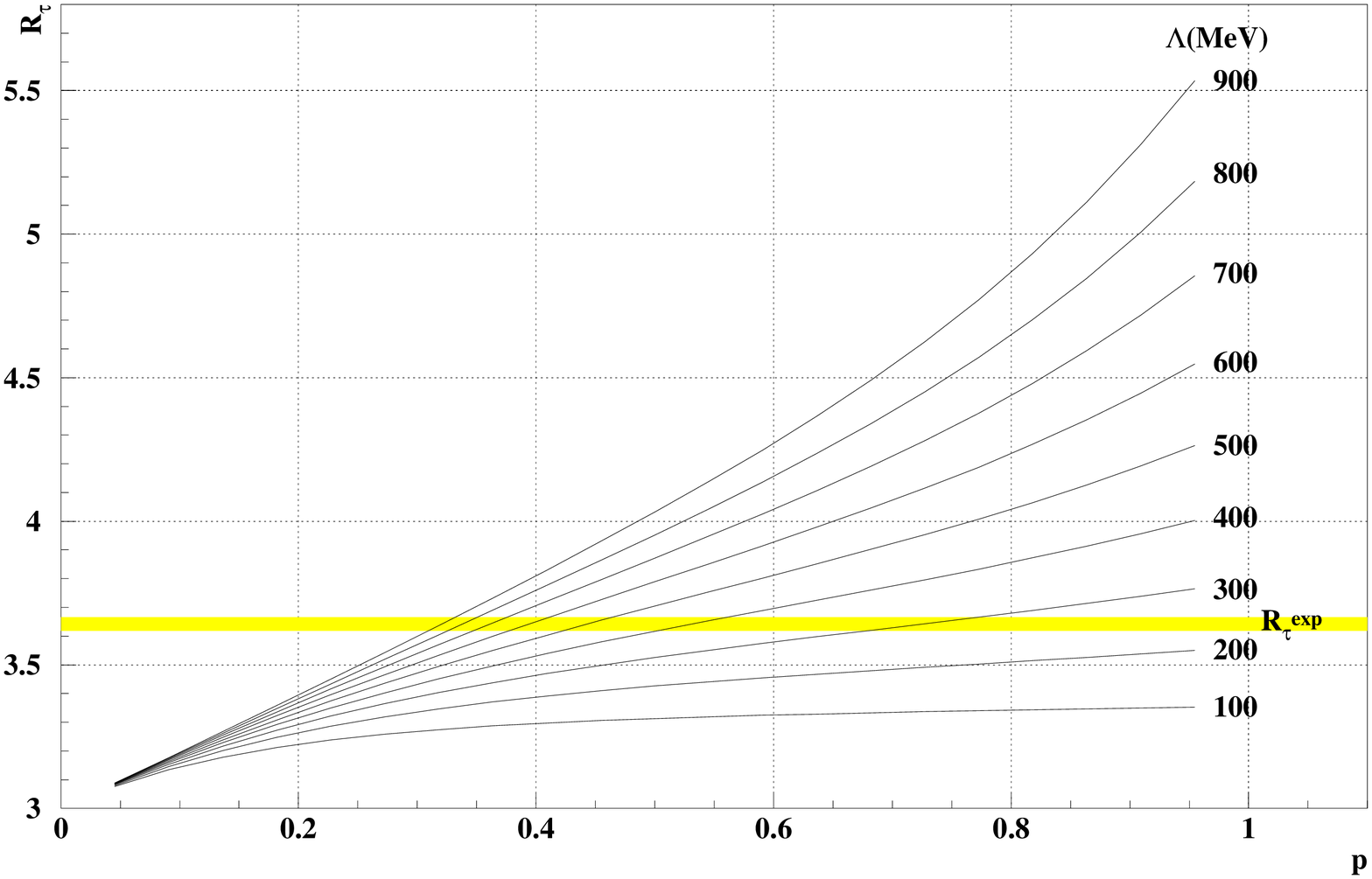}
\begin{minipage}[t]{10cm}\vspace{-.5cm}
Figure 4: $R_\tau$, obtained with different values of $p$ and 
$\Lambda^2$, compared with its experimental value (gray band). 
\end{minipage}
\end{figure*}

\section { Instability of the perturbative QCD vacuum}

In the previous section, we imposed confinement along with
AF. Here we discuss further results of some importance regarding
the nature of confinement itself through a vacuum instability 
induced by AF. The 1-loop result for $\im\big(\epsilon(s)\big)$ may
be decomposed into its quark and glue pieces as
$$
\im\big(\epsilon(s)\big) = - \pi b_{TOT} \vartheta(s), \eqno(12)
$$
with
$$
b_{TOT} = b_q + b_g > 0 \eqno(13)
$$
(for $N_F<16$), $b_q <0$ and $b_g >0$.\\
In QED on the other hand, only  charged particles contribute and
the sign in Eq.(12) is reversed. (The ``price'' for it is the Landau 
ghost absent in QCD).\par
In QED, this positive sign is necessary for stability, since
$\epsilon$ is related to the conductivity $\sigma$
$$
\epsilon(s) = \epsilon_o + {{i\sigma(s)}\over{\sqrt{s}}}.\eqno(14)
$$ 
Thus, for example, for Ohm's law to work, there must be dissipation,
i.e.,  $\re \big( \sigma(s)\big) > 0$. For such ``normal'' systems, 
it is usual to define a noise temperature $T_n$ which is positive.
\par
What if the sign is reversed ? Then the system is not dissipative
instead is an ``amplifier'' for which, the noise temperature $T_n<0$.
The notion of a noise temperature may be appreciated by considering
a system with two energy levels $E_1>E_0$. 
The probability ratio for finding the state with these energies
is given by
$$
{{P_1}\over{P_0}} = e^{-{{(E_1 - E_0)}\over{k_BT_n}}},\eqno(15)
$$
Thus, we see that the ``normal'' situation is for $T_n>0$ whereas the
``amplifier'' case has $T_n<0$. An artificially pumped system 
such as a MASER or a spin system has $T_n<0$. But such a system 
is {\bf unstable}. We have shown elsewhere that such systems exhibit
a Klein paradox for photons\cite{Panella}.\par
Turning to QCD then, since the perturbative ground state has
$\im\big(\epsilon(s)\big) < 0$, we conclude that such a system is
unstable. This is a pleasing physical result since it
implies that the perturbative ground state containing free quarks and
glue is an excited (higher energy) state whereas the states of lower energy 
containing the hadrons must be the true ground state of QCD.

\section{acknowledgments}
We acknowledge partial support from EEC contract TMR-CT98-0169.

\end{document}